\documentclass[useAMS,usenatbib]{mn2e}
\usepackage{graphicx}
\usepackage{lscape}
\usepackage{color}

\newcommand{\todcor}{\textsc{todcor} }

\newcommand{\jkt}{\textsc{jktebop} }
\newcommand{\jktabs}{\textsc{jktabsdim} }

\newcommand{\ms}{m~s$^{-1}$}

\title[ASAS~J195222-3233.7]{Orbital and physical parameters of eclipsing binaries from the ASAS 
catalogue --- VIII. The totally-eclipsing double-giant system HD~187669\thanks{Based on observations collected at the European Southern Observatory, Chile under programmes 085.C-0614, 085.D-0395, 086.D-0078, 087.C-0012, 089.C-0415, 190.D-0237, and 091.D-0469.}}
\author[K. G. He\l miniak et al.]{K. G. He\l miniak$^{1,2}$\thanks{Subaru Research Fellow; e-mail: xysiek@naoj.org}
D. Graczyk$^{3,4}$\thanks{e-mail: darek@astro-udec.cl},
M. Konacki$^{2,5}$,
B. Pilecki$^{6,4}$,
M. Ratajczak$^2$,
\newauthor
G. Pietrzy\'nski$^{6,4}$,
P. Sybilski$^2$,
S. Villanova$^4$,
W. Gieren$^{4,3}$,
G. Pojma\'nski$^6$, 
\newauthor
P. Konorski$^6$,
K. Suchomska$^6$,
D. E. Reichart$^7$,
K. M. Ivarsen$^7$,
J. B. Haislip$^7$,
\newauthor
and A. P. LaCluyze$^7$\\
$^{1}$Subaru Telescope, National Astronomical Observatory of Japan, 650 North Aohoku Place, Hilo, HI 96720, USA\\
$^{2}$Nicolaus Copernicus Astronomical Center, Department of Astrophysics, ul. Rabia\'nska 8, 87-100 Toru\'n, Poland\\
$^{3}$Millenium Institute of Astrophysics, Av. Vicu\~{n}a Mackenna 4860, Santiago, Chile\\
$^{4}$Universidad de Concepci\'on, Departamento de Astronom\'ia, Casilla 160-C, Concepci\'on, Chile\\
$^{5}$Astronomical Observatory, A. Mickiewicz University, ul. S\l oneczna 36, 60-286 Pozna\'n, Poland\\
$^{6}$Warsaw University Observatory, Al. Ujazdowskie 4, 00-478 Warsaw, Poland\\
$^{7}$Department of Physics and Astronomy, University of North Carolina, Campus Boc 3255, Chapel Hill, NC 27599-3255
}
\begin{document}

\date{Accepted ... . Received ...; in original form ...}

\pagerange{\pageref{firstpage}--\pageref{lastpage}} \pubyear{2015}

\maketitle

\label{firstpage}

\begin{abstract}
We present the first full orbital and physical analysis of HD~187669, 
recognized by the All-Sky Automated Survey (ASAS) as the eclipsing binary 
ASAS~J195222-3233.7. We combined multi-band photometry from the
ASAS and SuperWASP public archives and 0.41-m PROMPT robotic telescopes with our
high-precision radial velocities from the HARPS spectrograph. Two different 
approaches were used for the analysis: 1)~fitting to all data simultaneously 
with the WD code, and 2) analysing each light curve (with \textsc{jktebop}) and RVs 
separately and combining the partial results at the end.
This system also shows a total primary (deeper) eclipse, lasting for about 6 days.
A spectrum obtained during this eclipse was used to perform atmospheric analysis
with the \textsc{moog} and \textsc{sme} codes in order to constrain physical parameters of the
secondary.

We found that ASAS~J195222-3233.7 is a double-lined spectroscopic binary 
composed of two evolved, late-type giants, with masses of $M_1 = 1.504\pm0.004$ 
and $M_2=1.505\pm0.004$~M$_\odot$, and radii of $R_1 = 11.33\pm0.28$ and 
$R_2=22.62\pm0.50$~R$_\odot$, slightly less metal abundant than the Sun, 
on a $P=88.39$~d orbit.
Its properties are well reproduced by a 2.38 Gyr isochrone, 
and thanks to the metallicity estimation from the totality spectrum and high 
precision in masses, it was possible to constrain the age down to 0.1 Gyr.
It is the first so evolved galactic eclipsing binary measured with such 
a good accuracy, and as such is a unique benchmark for studying the 
late stages of stellar evolution.
\end{abstract}


\section{Introduction}

Despite the fortunate configuration of detached eclipsing binaries (DEBs)
and many possibilities that it gives us, analysis of these objects still
encounters some difficulties. The light curves themselves do not contain
enough information about the effective temperatures in the absolute scale, 
mainly about their ratio. It is sometimes being set on the basis of the colour
of the whole system, so combined light of two, sometimes very different stars.
Another problem occurs when it comes to calculate the fractional radii 
(defined as a fraction of the major semi-axis). The information about their sum
comes mainly from the width of eclipses, and is somewhat degenerated with the 
inclination angle, but from the light curves only it is difficult to constrain 
their ratio. Again, other kinds of data are needed, like spectra, from which one 
can try to estimate the ratio of fluxes coming from the two components. 
Both that issues are however much less important in even more fortunate case,
when a system shows total eclipses, when light from only one component 
is seen. The presence of a flat 
minimum in the light curves already solves the mentioned problems
and other kind of observation help to improve the analysis even more.

Such a fortunate situation occurs either when the inclination angle is very
close to 90 degrees, or when the two stars have significantly different sizes.
The latter usually means that at least one component is evolved. Because of a 
long-lasting evolution on the main sequence (MS), such evolved systems are much less 
common than the MS eclipsing binaries. In the recent, very fine summary
\citet{tor10} point out the lack of red giant systems with accurately measured
properties, especially masses and radii. \citet{tor10} list only 4 red giants
in their sample: AI~Phe~A, TZ~For~A, and both components of 
OGLE-051019.64-685812.3 in the LMC. Since then a small number of systems
have been added to the sample, but either containing one giant component
\citep[KIC~8410637;][]{fra13}, or located the Magellanic Clouds
\citep[e.g.][]{pie13,gra14}, i.e. no galactic double-giant system has 
been accurately studied. Some interesting cases were analysed 
\citep{gal08,rat13} but due to various reasons their parameters are not yet determined
precisely enough. Long baseline interferometry was successful in measuring
the radii of single red giants directly, but without mass determination. 
Asteroseismology of solar-type oscillations is another option, and with long-cadence,
continuous and precise light curves from {\it CoRoT} and {\it Kepler} satellites it
appears to be a promising method \citep{kal09,bed10}, especially if combined
with interferometric radius measurements \citep{bai14}, but still the precision
achieved is lower than for double-lined DEBs, or the differences between
parameters obtained from asteroseismology and other methods is significant.

In this paper we present our results of a detailed analysis of a binary system
showing a total eclipse, and composed of two cool giant stars -- ASAS~J195222-3233.7 
(HD~187669, CD-32~15534, TYC~7443-867-1; hereafter ASAS-19). 
Despite being relatively bright -- $V\sim8.9$~mag -- this star was recognized as 
a binary only in the All-Sky Automated Survey 
\citep[ASAS;][]{poj02}\footnote{\texttt{http://www.astrouw.edu.pl/asas/?page=acvs}} 
and this is the first detailed study of this interesting target. 
Time-series photometry is also available in the Public Archive of the
Wide-Angle Search for Planets \citep[SuperWASP;][]{pol06}. Except single-epoch 
brightness and position measurements, no information is available in other 
data bases or literature. The only spectral type classification -- K0III --
comes from \citet{hou82}. 

Two teams were working on this system mostly independently. One group was led by K. He{\l}miniak (H-group, including MK, MR, PS) and second group by D. Graczyk (G-group, including BP, GP, PK, SV, WG, KS). 
We used the same data in our analysis and we consulted our partial results as the work progressed. However, overall approach used by each group was different. In the end we combined our results to obtain the final parameters of the system.

\section{Observations}

\begin{table}
\centering
\caption{The PROMPT $V,I$ and ASAS $I$ photometry of ASAS-19.
Portion of the table is shown for the reference.
The complete table is available in the on-line version of the
manuscript.}\label{tab_lc_online}
\begin{tabular}{lccc}
\hline\hline
BJD-2450000 & Mag & err & Set \\
\hline
2404.77762&	7.567&	0.075&	AI\\
2405.80652&	7.480&	0.071&	AI\\
2406.82007&	7.502&	0.074&	AI\\
2415.82223&	7.494&	0.068&	AI\\
2500.62185&	7.533&	0.074&	AI\\
...&&&\\
\hline
\end{tabular}
\end{table}

\begin{table*}
\centering
\caption{RV measurements from disentangling and least-squares spectra 
fitting (H-group), and RaVeSpAn (G-group), and their residuals (all in k\ms). Index ``1''
denotes the hotter star (primary) and ``2'' the cooler (secondary)}. 
\label{tab_rv_kh}
\begin{tabular}{@{}lrcrclrcrc}
\hline \hline
 & \multicolumn{4}{c}{\bf H-group} & & \multicolumn{4}{c}{\bf G-group} \\
JD-2450000 & $v_1$ & $(O\!-\!C)_1$ & $v_2$ & $(O\!-\!C)_2$ & & $v_1$ & $(O\!-\!C)_1$ & $v_2$ & $(O\!-\!C)_2$ \\
\hline
5432.55909 &  17.346 &  0.001 & -48.742 &  0.057 & &  16.954 &  0.014 & -49.149 &  0.075 \\
5467.50740 & -47.938 & -0.010 &  16.282 & -0.071 & & -48.140 &  0.016 &  15.823 & -0.019 \\
5468.49296 & -48.725 &  0.017 &  17.104 & -0.062 & & -49.002 & -0.035 &  16.582 & -0.069 \\
5470.48528 & -49.883 &  0.008 &  18.228 & -0.086 & & -50.101 &  0.011 &  17.705 & -0.088 \\
5471.48792 & -50.220 & -0.007 &  18.589 & -0.046 & & -50.443 & -0.011 &  18.086 & -0.027 \\
5477.66139 & -48.393 & -0.037 &  16.770 & -0.010 & & -48.645 & -0.075 &  16.261 &  0.000 \\
5478.66494 & -47.458 & -0.013 &  15.852 & -0.019 & & -47.663 & -0.004 &  15.345 & -0.009 \\
5479.50341 & -46.618 & -0.057 &  14.973 & -0.015 & & -46.829 & -0.054 &  14.495 &  0.022 \\
5503.51201 &   3.903 & -0.030 & -35.459 & -0.041 & &   3.570 & -0.056 & -35.968 & -0.024 \\
5504.50635 &   5.859 & -0.024 & -37.428 & -0.066 & &   5.546 & -0.023 & -37.940 & -0.055 \\
5721.64681 & -29.487 &  0.073 &  -1.956 &  0.031 & & -29.733 &  0.140 &  -2.448 & -0.022 \\
5721.75742 & -29.746 &  0.063 &  -1.700 &  0.038 & & -30.050 &  0.073 &  -2.150 &  0.026 \\
5722.65625 & -31.774 &  0.022 &   0.286 &  0.039 & & -32.106 & -0.005 &  -0.199 &  0.000 \\
5722.77460 & -32.030 &  0.023 &   0.557 &  0.054 & & -32.389 & -0.032 &   0.095 &  0.037 \\
5811.58372 & -32.927 &  0.029 &   1.471 &  0.067 & & -33.268 & -0.006 &   0.972 &  0.009 \\
5813.59359 & -37.061 & -0.010 &   5.645 &  0.152 & & -37.374 & -0.032 &   5.103 &  0.063 \\
6137.54467 &  18.418 &  0.020 & -49.839 &  0.011 & &  17.970 & -0.021 & -50.208 &  0.065 \\
6138.52970 &  18.009 &  0.002 & -49.492 & -0.032 & &  17.557 & -0.036 & -49.896 & -0.021 \\
6178.63170 & -50.258 & -0.038 &  18.669 &  0.027 & & -50.462 & -0.018 &  18.151 &  0.026 \\
6178.69965 & -50.229 &  0.006 &  18.687 &  0.030 & & -50.472 & -0.014 &  18.183 &  0.043 \\
6179.54739 & -50.351 &  0.006 &  18.768 & -0.011 & & -50.556 &  0.019 &  18.274 &  0.018 \\
6179.66679 & -50.354 &  0.010 &  18.782 & -0.004 & & -50.545 &  0.037 &  18.295 &  0.032 \\
6179.69162 & -50.374 & -0.009 &  18.788 &  0.002 & & -50.603 & -0.020 &  18.300 &  0.036 \\
6214.49823 &  10.895 &  0.013 & -42.425 & -0.074 & &  10.573 & -0.013 & -42.901 & -0.007 \\
6240.54579 &  -2.315 & -0.054 & -29.200 &  0.035 & &  -2.614 &  0.088 & -29.579 &  0.020 \\
6448.94701 & -49.097 & -0.002 &  17.506 & -0.013 & & -49.285 &  0.003 &  16.964 & -0.012 \\
\hline
\end{tabular}
\end{table*}

\subsection{Photometry}
\subsubsection{ASAS}
The $V$-band photometry of ASAS-19, publicly available from the ASAS 
Catalogue\footnote{\texttt{http://www.astrouw.edu.pl/asas/?page=aasc}},
spans from November 2000 to December 2009, and contains 406 good quality 
points (flagged ``A'' in the original data). 

The $I$-band photometry was downloaded from internal ASAS catalogue and spans from
May 2000 to June 2009, and contains 247 good points.

\subsubsection{SuperWASP}
From the SuperWASP public 
archive\footnote{\texttt{http://exoplanetarchive.ipac.caltech.edu/applications/
/ExoTables/search.html?dataset=superwasptimeseries}}
we have extracted raw flux measurements of the binary. In order to transform
them to magnitudes, we used flux measurements of a nearby, slightly brighter
star HD~187742 ($V=8.07$~mag, $SW=8.193$~mag), also classified as K0III
\citep{hou82}, which we have previously inspected for variability.
We cross-matched the two data sets and removed the obvious outliers from the 
resulting light curve. Originally, the SuperWASP data spanned from 
March 2006 to May 2008 (three observing seasons), but we have found that 2007 
and 2008 data suffer from large systematic variations, thus we decided to 
include data only from April and July 2007, when the primary eclipse was 
recorded, and the observations do not outlay significantly.
We ended up with 5554 good data points.

\subsubsection{PROMPT}
Dedicated photometric observations of ASAS-19 were carried out 
in $V$ and $I$ bands with the 0.41-m Prompt-4 and Prompt-5 robotic 
telescopes\footnote{Panchromatic Robotic Optical Monitoring and 
Polarimetry Telescopes. PROMPT is operated by SKYNET -- a distributed 
network of robotic telescopes located around the World, dedicated for 
continues GRB afterglows observations. \texttt{http://skynet.unc.edu}}, 
located in the Cerro Tololo Inter-American Observatory in Chile. 
A more detailed description of the observational settings, reduction 
procedure and calibration to standard photometric system can be found in 
\citet{hel11}. The PROMPT observations span about 400 days. In total we secured 
1714 and 1400 measurements in $V$ and $I$ bands respectively.
The typical exposure times were 5-7 sec for $V$ and 2-3 sec for the $I$ band.
Most of the observations were concentrated in the two eclipses, especially
in the flat part of the primary one, covered almost completely in both
bands between September 20 and 25, 2009.

{Table \ref{tab_lc_online}} contains PROMPT $V$ and $I$-band, and ASAS $I$-band 
light curves. The first column is the time stamp BJD-2450000, second and third 
colums are the measured brightness (in mag) and its formal error. The last column 
denotes the data set: AI = ASAS $I$, PI = PROMPT $I$, and PV = PROMPT $V$. 
The complete table is available in the {machine-readable form in the
electronic version of the manuscript}.

\subsection{HARPS Spectroscopy}

ASAS-19 was observed spectroscopically with the High Accuracy 
Radial velocity Planet Searcher \citep[HARPS;][]{may03}, attached to the
3.6-m telescope in La Silla observatory, Chile, between August 2010 and 
June 2013. A total of 27 spectra were taken in two modes -- high efficiency
(EGGS) and high RV accuracy.

Fourteen spectra, taken between 2009 and 2013, were obtained 
in the efficiency EGGS mode. The exposure time was usually between 300 and 
600 seconds depending on a seeing conditions at La Silla.
We'd like to call special attention to the spectrum from September 10, 2010, 
taken exactly during the total part of the primary eclipse, when light from
only one component was recorded. This spectrum was used for atmospheric
analysis, but the radial velocity was not measured.

Thirteen spectra, taken between June 2011 and September 2012, were obtained in 
the high RV accuracy mode. The exposure time for those observations varied 
between 780 and 1200 seconds, giving the S/N around 5500\AA\ of 70-120. 
All spectra were reduced on-site with the available Data Reduction Software (DRS).

\section{Analysis}

\subsection{Radial velocities}

\begin{figure}
\includegraphics[width=\columnwidth]{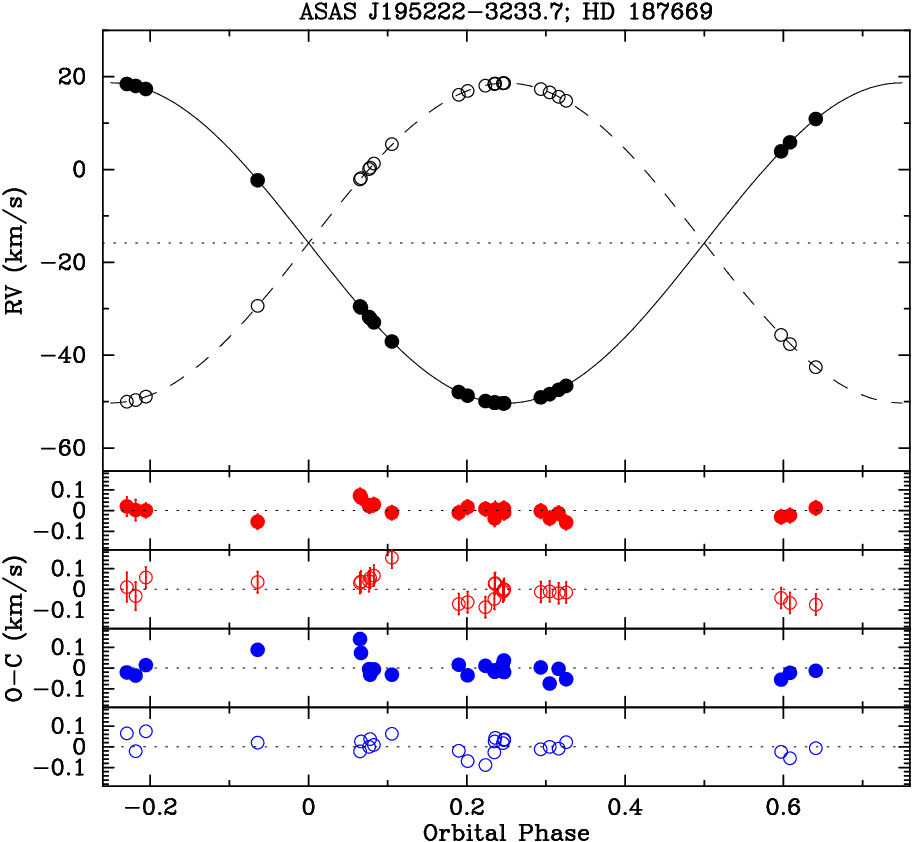}
\caption{RV measurements and best-fitting orbital solution for ASAS-19.
Solid line and filled symbols refer to the primary, and dashed line and open 
symbols to the secondary component. Differences in RV measurements
by two groups are smaller than the size of symbols, and models are 
practically indistinguishable.
Dotted line marks the systemic velocity of the primary.
The difference between the two systemic velocities was accounted 
for. Lower panels depict the residuals for each component and each group
(different fits) separately. Phase zero is set to the primary minimum. 
The resulting $rms$ are 30 and 54 \ms\ for the primary and secondary
respectively from the H-group's solution (red), and 43 and 42 \ms\
analogously for the G-group (blue).
Colour version of the figure available in the on-line version 
of the manuscript.
}\label{fig_rv_kh}
\end{figure}

\begin{table}
\centering
\caption{Results of the orbital fit to the RVs performed by the H-group.}
\label{tab_par_orb_kh}
\begin{tabular}{lcc}
\hline \hline
Parameter & Value & $\pm$ \\
\hline
$P$ (d) & 88.3891 & 0.0008 \\
$T_Q$ (JD)$^a$ & 2452069.851 & 0.043 \\
$K_1$ (k\ms) & 34.524 & 0.010 \\
$K_2$ (k\ms) & 34.461 & 0.015 \\
$\gamma_1$ (k\ms) & -15.846 & 0.008 \\
$\gamma_2-\gamma_1$ (k\ms) & 0.177 & 0.015 \\
$a_{12} \sin{i}$ (R$_\odot$) & 120.549 & 0.036\\
$e$ & 0.0 & (fix) \\
$q$ & 1.0018 & 0.0005 \\
$M_1 \sin^3{i}$ (M$_\odot$) & 1.5020 & 0.0013 \\
$M_2 \sin^3{i}$ (M$_\odot$) & 1.5047 & 0.0011 \\
$rms_1$ (\ms) & \multicolumn{2}{c}{30}\\
$rms_2$ (\ms) & \multicolumn{2}{c}{54}\\
DOF & \multicolumn{2}{c}{43} \\
$\chi^2/DOF$  & \multicolumn{2}{c}{0.9963}\\
\hline
\end{tabular}
\\$^a$ For a quadrature before the primary eclipse. \\Not adopted in further analysis.
\end{table}

\subsubsection{H-group}
Radial velocities (RVs) were initially calculated with the two-dimensional 
cross-correlation \todcor code \citep{zuc94}, with synthetic spectra taken 
as templates. These RVs were then used as starting values for the tomographic
spectral disentangling and least-squares fitting procedure 
\citep{kon10}. This procedure uses tomographic methods to produce
decomposed spectra of each star, suitable for more precise RV measurements and 
spectral analysis (after proper scaling). To find the new RVs, the code uses the 
least-squares method to find shifts of the two spectra in the $\log{\lambda}$ domain, 
so their sum matches a given observed spectrum.

\subsubsection{G-group}
Determination of components' radial velocities was done using RaVeSpAn code \citep{pil12} utilizing the Broadening Function formalism \citep{ruc92,ruc99}. We used templates from synthetic library of LTE spectra by \cite{col05}; the templates were not convolved down to the HARPS resolution. In the beginning we choose templates to match components' effective temperature, gravity and abundance. However the resulting {\it rms} of both radial velocity curves was significantly larger than those from H-group. We decided to investigate the effect. It turned out that using solar metallicity and cooler template (T$_{\rm eff}\approx 4000$ K) for both components resulted in reducing {\it rms} by a factor of 1.5. For more the difference in {\it rms} between both stars were reduced to almost zero signifying similar precision of their radial velocity determination. We could expect this because although the secondary rotates two times faster than the primary (producing larger rotational broadening of lines) at the same time it is optically 2.5 times brighter (producing significantly stronger lines in combined spectrum). Both effects should cancel out if there are not other important sources of the scatter (i.e. stellar spots). The resulting radial velocities have slightly larger rms than those derived by tomographic spectral disentangling. Also $\gamma$ difference between components is much smaller -- 40 \ms -- and comparable with individual $rms$ (see further Sections). The overall precision of RV measurements and orbital solutions made by both groups is slightly worse than expected from the spectrograph's performance. It is probably because of a noticeable rotational broadening of both components and/or stellar activity. Our measurements and their residuals from the WD fit are shown in the {Table~\ref{tab_rv_kh}}.

\subsection{Spectroscopic orbital fit (H-group)}
The strategy of the H-group was to obtain partial results with different
approaches and working on different data, and combine them into one set later.
The orbital fit to the RVs measured by least-squares fitting was done first.
The fit was performed with the \textsc{v2fit} code, which is a simple 
procedure that fits a double-keplerian solution with a Levenberg-Marquardt 
algorithm. As free parameters we set the two velocity semi-amplitudes $K_{1,2}$,
orbital period $P$, centre-of-mass velocity of the primary $\gamma_1$, 
difference between two centre-of-mass velocities $\gamma_2-\gamma_1$,
and a time of phase zero, defined as moment of the periastron passage
for eccentric orbits, or a quadrature for circular orbits. Initially we 
also set free the eccentricity $e$ and argument of the periastron
$\omega$, but we have found $e$ to be indifferent from zero. 

We have found however that the two components have significantly 
different values of $\gamma$, with the primary's (defined here as 
the hotter star) being blueshifted by $177\pm15$~\ms~ -- larger
than the G-group. 
Several explanations are possible, but the one we find the most plausible
is that it is a systematic introduced by the method used by the H-group, 
which is optimized for precise measurements of velocity variations, not their
absolute values. We also find possible that it was due to stellar 
spots, which caused time-varying asymmetries in
line profiles, which finally led to a template mismatch,
or due to different large-scale convective motions in
the two stars \citep{sch75,por00}.
We can exclude the differential gravitational redshift, as 
it would make the secondary blueshifted.

The measurement errors of the order of single \ms\ occur to be underestimated, 
so to get the reduced $\chi^2$ close to 1, thus reliable statistical errors of 
the parameters, we added in quadrature a systematic contribution of of 36 and 52 \ms\ 
for the primary and secondary respectively. To account for possible 
systematics in the final solution, we run 10000 Monte Carlo iterations, 
perturbing the parameters that were held fixed (i.e. $e$ and $\omega$).
We added the MC errors to the statistical ones in quadrature, however they
were typically an order of magnitude lower than the statistical ones. 
All the RV measurements from the tomographic disentangling, together with their 
residuals from the model RV curve, are shown in the Table~\ref{tab_rv_kh}.
Neither of the groups used the spectrum taken in totality for the RV
calculations and further modelling.
The resulting orbital parameters are presented in Table~\ref{tab_par_orb_kh},
and the corresponding model RV curves are shown in Figure~\ref{fig_rv_kh}.

\begin{figure*}
\includegraphics[width=0.7\textwidth]{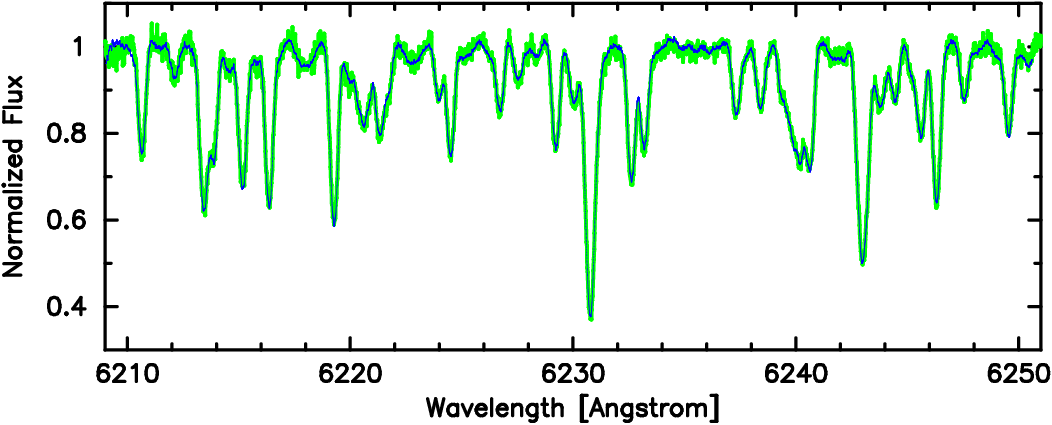}
\caption{Comparison of the spectrum recorded during the total eclipse (green)
with the rescaled secondary's spectrum from the disentangling (blue). The match is
almost perfect, but the disentangled spectrum is of much higher S/N.
Colour version of the figure available in the on-line version 
of the manuscript.
}\label{fig_tomo}
\end{figure*}

\begin{figure*}
\includegraphics[width=0.8\textwidth]{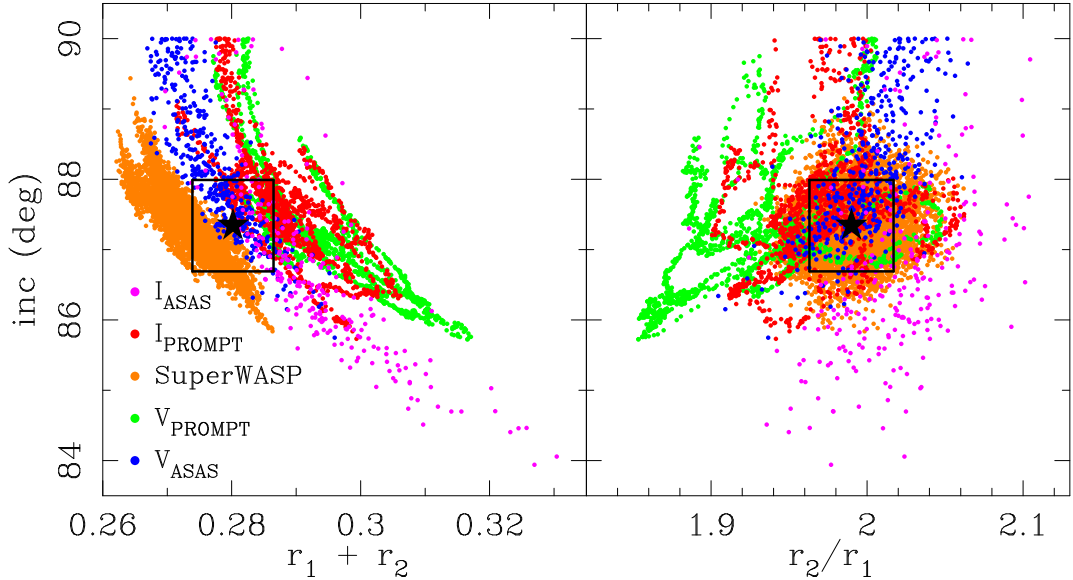}
\caption{Results of the residual-shifts analysis performed with \jkt on all the
data sets separately. Plots present the distribution of consecutive solutions on the
$r_1 + r_2$ vs. $i$ (left) and $k=r_2/r_1$ vs. $i$ (right) panels. Black stars and boxes
correspond to the adopted values with their $1\sigma$ uncertainties. Correlation 
between $r_1 + r_2$ and the inclination is clear, however the inclination itself 
does not change significantly from set to set.
Colour version of the figure available in the on-line version of the manuscript.
}\label{fig_inc_rr}
\end{figure*}

\subsection{Spectral analysis of the decomposed and total eclipse spectra}

\subsubsection{\textsc{moog} (G-group)} 
We disentangled spectra of both components and then we analysed them together with the single spectrum of the secondary component taken at the total primary eclipse. As for disentangling and atmospheric parameters derivation we used the LTE program \textsc{moog} \citep{sne73} and follow the prescription given in \citet{gra14}. Details of the method are given in \cite{mar08} and the line list in \cite{vil10}. The totality spectrum was analysed first, and the temperature $T_{{\rm eff},2}=4360$~K was obtained. The disentangled spectra were scaled using the light ratio determined from solution of radial velocity and light curves, assuming temperature $T_{{\rm eff},2}=4360$~K, by fitting e.g. temperature of the primary $T_{{\rm eff},1}$. The light ratio varied from 2.2490 at 4670~\AA, to 2.6712 at 6470~\AA. The results are summarized in Tab.~\ref{spec_res}. Typical errors in $T_{\rm eff}$, $\log{g}$, [$Fe/H$] and $v_t$ are 70 K, 0.3, 0.15 dex and 0.2 km s$^{-1}$, respectively. Regarding uncertainties parameters derived from the totality spectrum are consistent with those obtained from the disentangled spectrum of the secondary. The small differences on a level of 1$\sigma$ are caused by a little larger depth of absorption lines in disentangled spectrum. 

The same procedure of deriving $T_{\rm eff}$ as used here (methodology, and data from HARPS), for Arcturus, a standard star as far as $T_{\rm eff}$ is concern, gives 4290 K \citep{vil10}, which agrees very well with independent measurements \citep[e.g.][which gives $T_{\rm eff}=4286$ K]{ram11}. So, in spite of using an LTE approximation, 
we can recover a reliable $T_{\rm eff}$ for such kind of stars (cold giants at that metallicity) which is essentially free from larger systematic errors.

\begin{table}
\centering
\caption{Atmospheric parameters from the \textsc{moog} (G-group)}
\label{spec_res}
\begin{tabular}{lcccc}
\hline \hline
Spectrum& $T_{\rm eff}$&$\log{g} $& [$Fe/H$]& $v_t$ \\
   & (K) & (cgs) & (dex) & (km s$^{-1}$)\\
\hline
primary& 4770& 2.30 &$-0.25$& 1.25\\
secondary&4440&1.60 &$-0.22$& 1.61\\
totality& 4360& 1.57 &$-0.44$&1.65\\
adopted$^a$ & 4360 & 1.90$^b$& $-0.30 $ & 1.65 \\  
\hline
\end{tabular}
\\
$^a$ For the secondary.\\
$^b$ From the WD solution.
\end{table}

\begin{table}
\centering
\caption{Atmospheric parameters from the \textsc{sme} (H-group)}
\label{tab_sme}
\begin{tabular}{lccc}
\hline \hline
Spectrum& $T_{\rm eff}$& [$Fe/H$] & $v_{\rm rot}$ \\
   & (K) & (dex) & (km s$^{-1}$)\\
\hline
primary&  4610& $-0.24$&  6.87\\
secondary&4310& $-0.21$& 13.60\\
totality& 4290& $-0.19$& 13.56\\
adopted$^a$ & 4300 & $-0.20 $ & 13.58 \\  
\hline
\end{tabular}
\\
$^a$ For the secondary, from 10 runs.
\end{table}

\subsubsection{\textsc{sme} (H-group)}
We also analysed the disentangled and total eclipse spectra 
with the Spectroscopy Made Easy \citep[\textsc{sme};][]{val96}. 
To ensure that the disentangled spectra are properly scaled, we have 
used the flux ratios obtained for each echelle order separately, 
taken from our initial \todcor measurements. In the range of the $V$
band they were in a good agreement with the flux ratio obtained from 
the \jkt solution (next Section). We have also compared the scaled 
disentangled spectrum of the secondary with the spectrum in totality, 
and found almost perfect match (Fig. \ref{fig_tomo}).
 
We run the \textsc{sme} separately on five HARPS orders between 5907 
and 6215 \AA, with $\log(g)$ being kept fixed to 2.507 and 1.907 
for the primary and secondary, respectively -- values found in the 
analysis described in further Sections. For a given component, 
all runs gave consistent values of $T_{\rm eff}$, [$Fe/H$] and $v_{rot}$, 
the last one being in agreement with the results expected from the 
measured {radii}, assuming spin-orbit alignment and 
rotational synchronization. 
As final results we adopted average values of all five runs for the 
primary, and ten (disentangled + totality) for the secondary, and standard 
deviations as their uncertainties. {We got $T_{\rm eff, 1}=4610\pm50$~K, 
$[Fe/H]_1=-0.24\pm0.12$~dex, $T_{\rm eff,2}=4300\pm50$~K, and $[Fe/H]_2=-0.20\pm0.07$~dex. 
Except $T_{\rm eff, 1}$, all values are in a better than 1$\sigma$ agreement with the
ones adopted by the G-group (Table \ref{spec_res}). However, the final
value of $T_{\rm eff, 1}$ by the G-group is somewhat lower (Sect. \ref{sec_wd}),
and also consistent within 1$\sigma$ with our \textsc{sme} analysis.}
We summarise our \textsc{sme} results in Table~\ref{tab_sme}. 
Uncertainties of $v_{\rm rot}$ are 0.3 km s$^{-1}$.

Additionally, we estimated the secondary's effective temperature from
the $V-I$ colour vs. line-depth ratio calibrations by \citet{str00}. 
We used the totality spectrum and measured 10 ratios of 
metallic lines from the 6380-6460 \AA\ region, and got the intrinsic 
secondary's colour $(V_2\!-\!I_2)_0=1.228\pm0.030$~mag. 
This corresponds to $T_{\rm eff,2}=4370\pm80$~K \citep{wor11}, and a 
K2.5-3~III star \citep{tok00}.

\subsection{Light curve solution with JKTEBOP (H-group)}\label{sec_jkt_h}

\begin{table*}
\centering
\caption{Results of the \jkt fit to the observed LCs (H-group).}
\label{tab_lc_jkt}
\begin{tabular}{lcccccc}
\hline \hline
Parameter & $I_{ASAS}$ & $V_{ASAS}$ & $I_{PROMPT}$ & $V_{PROMPT}$ & SuperWASP & Adopted \\
\hline
$T_0$ (JD-2452000)$^a$ & 92.118(37) & 92.036(47) & 92.085(28) & 92.095(30) & 92.058(17) & 92.074(25) \\
$r_1+r_2$	& 0.2929(95) & 0.2769(57) & 0.2875(73) & 0.2931(98) & 0.2736(46) & 0.2802(63) \\
$k = r_2/r_1$	&  2.014(44) &  2.002(23) &  1.975(29) &  1.933(44) &  1.992(22) &  1.990(27) \\
$i$ (deg)	&  86.5(1.2) &  88.0(1.1) &  87.52(62) &  87.20(79) &  87.30(46) &  87.34(65) \\
$r_1$		& 0.0971(40) & 0.0923(24) & 0.0967(31) & 0.0999(49) & 0.0914(18) & 0.0937(23)$^b$ \\
$r_2$		& 0.1958(62) & 0.1847(34) & 0.1909(44) & 0.1932(54) & 0.1821(30) & 0.1865(59)$^b$ \\
$(L_2/L_1)_I$	&  2.934(93) &    ---     &  2.822(82) &    ---     &    ---     &  2.871(87) \\
$(L_2/L_1)_V$	&    ---     &  2.421(43) &    ---     &  2.413(78) &    ---     &  2.419(51) \\
$(L_2/L_1)_{SW}$&    ---     &    ---     &    ---     &    ---     &  2.404(27) &  2.404(27) \\
$rms$ (mag)	& 0.025 & 0.017 & 0.017 & 0.017 & 0.011 &  \\
\hline
\end{tabular}
\\$^a$ Mid-time of the primary eclipse; $^b$ From the adopted sum and ratio of radii.
\end{table*}

One of the codes we used for the light curve (LC) analysis was the 
version v28 of \jkt \citep{sou04a,sou04b}, which is based on the 
\textsc{ebop} program \citep{pop81}. It is a fast procedure working on 
one set of photometric data at a time, not allowing for analysis of RV 
curves. On the basis of spectroscopic data we first found the mass ratio 
and orbital period, which we included in the LC analysis. We found that the
orbital period found directly by \jkt is in agreement with the one from RVs, 
however leading to significantly worse orbital solution. It is because of a
longer time span of spectroscopy with respect to PROMPT and SuperWASP 
observations, and that ASAS data do not include many points in the eclipses.

For \jkt we used the logarithmic limb darkening (LD) law with coefficients
interpolated from the tables of \citet{vHa96} for ASAS and PROMPT. 
For the SuperWASP data we used tables calculated by the developers of the 
\textsc{phoebe} code\footnote{\texttt{http://phoebe-project.org/1.0/files/ld/swasp\_2006.ld}}.
The gravity darkening coefficients and bolometric albedos were 
always kept fixed at the values appropriate for stars with convective envelopes 
($g = 0.32$, $A = 0.5$). As mentioned before, no significant eccentricity of 
the orbit of ASAS-19 was found, nor the third light, thus $e$ and $L_3$ were 
kept fixed to 0. We fitted for the sum of the
fractional radii $r_1+r_2$, their ratio $k$, orbital inclination $i$,
moment of the primary minimum $T_0$, surface brightness ratios $J$, and 
brightness scales (out-of-eclipse magnitudes in each filter).

To calculate reliable errors, we run the task 9, which uses the residual-shifts 
method \citep{sou08} to asses the importance of the correlated `red' noise, 
especially strong in the SuperWASP data \citep{sou11}. 
We have run several tests to check how the final model varies with various 
LD coefficients and ephemeris, and we did not notice a strong dependence, 
but to at least partially account for LD coefficients and ephemeris uncertainties, 
we let them to be perturbed in the residual-shifts simulations. 
It is a known fact that orbital inclination is correlated with the 
radii-related parameters, especially their sum. 
In Figure~\ref{fig_inc_rr} we show the results of the \jkt analysis on the 
$r_1+r_2$ vs. $i$, and $k=r_2/r_1$ vs. $i$ diagrams. We see that different 
data sets give similar values of inclination and $k$,
but clearly different areas of the $r_1+r_2$ vs. $i$ plane are occupied. 
The most likely reason for this inconsistency is the the activity and the 
location of spots, probably varying in time, and which was not included in 
the \jkt analysis. As shown for late-type dwarfs \citep[for example:][]{roz09,win10,hel11}, 
location of spots on different components may lead to variations in resulting 
radii reaching 2-3 per cent, while the accuracy of our photometry may not be 
enough to detect the spot-originated brightness variations. 

\begin{figure}
\includegraphics[width=\columnwidth]{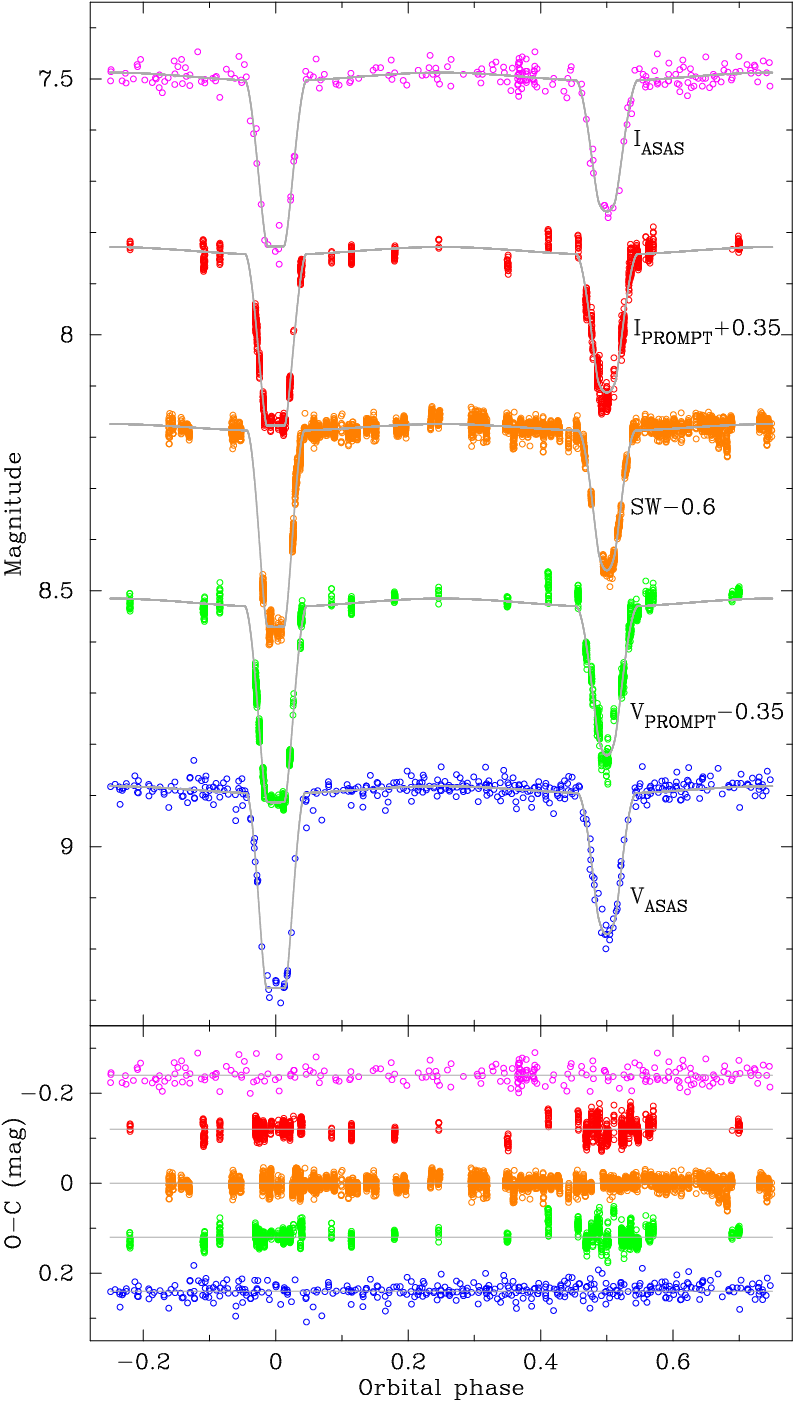}
\caption{{\it Top:} Photometry of ASAS-19 (open circles) and \jkt models 
(grey lines) for each band. PROMPT and SuperWASP data were shifted for clarity
by the indicated values (in mag). {\it Bottom:} Residuals of the \jkt models, 
shifted for clarity. Colour coding and sequence is the same as above.
Colour version of the figure available in the on-line version of the manuscript.
}\label{fig_lc_jkt}
\end{figure}

As the resulting parameters we adopted weighted averages of the values found
from the five data sets. We mark them in Figure~\ref{fig_inc_rr}, together
with the adopted 1$\sigma$ errors. The model LCs are presented in 
Fig.~\ref{fig_lc_jkt}. Looking at the scatter of the PROMPT photometry
in both eclipses, we can conclude that more spots reside on the primary 
(hotter, smaller) component. If so, the $rms$ of the H-group's RV 
measurements of both components is more likely enhanced by the 
rotational broadening, than the activity. If it was activity, 
we could expect larger $rms$ for the (slower rotating) primary, but
we observe the opposite. The resulting values of fractional radii 
$r_{1,2}$, and the inclination are given in Table~\ref{tab_lc_jkt}.
The oblateness of both components is below 1\%, so the usage of \jkt is justified.

Finally, we have used the \jkt solutions to derive observed $V-I$ colours 
of both components, and estimate their effective temperatures. Please note, 
that these simple calculations are possible only for totally-eclipsing systems.
For the secondary we have simply used the photometry in the total eclipse 
and got $1.434(1)$~mag. Taking the intrinsic $(V_2\!-\!I_2)_0=1.228(30)$~mag 
from line depth ratios, we get the value of $E(V\!-\!I)=0.206(30)$~mag, and 
$E(B\!-\!V)=0.161(23)$, assuming $E(V\!-\!I)=1.28\,E(B\!-\!V)$. 
From the light curve solutions we got magnitude differences between the 
components: $V_2\!-\!V_1=-0.959(23)$ and $I_2\!-\!I_1=-1.145(32)$~mag. 
We then get the observed primary's $V\!-\!I=1.082(39)$~mag, and its 
intrinsic value of $1.046(39)$~mag. This corresponds to 
$T_{\rm eff,1}=4710\pm110$~K, \citep{wor11} and a K0.5~III star \citep{tok00}. 
Interestingly, both temperatures obtained from the calibrations
of \citet{wor11} -- 4710 and 4370~K -- are 1.7 per cent larger than those 
from our \textsc{sme} analysis (4630 and 4300~K).

\begin{figure}
\includegraphics[width=\columnwidth]{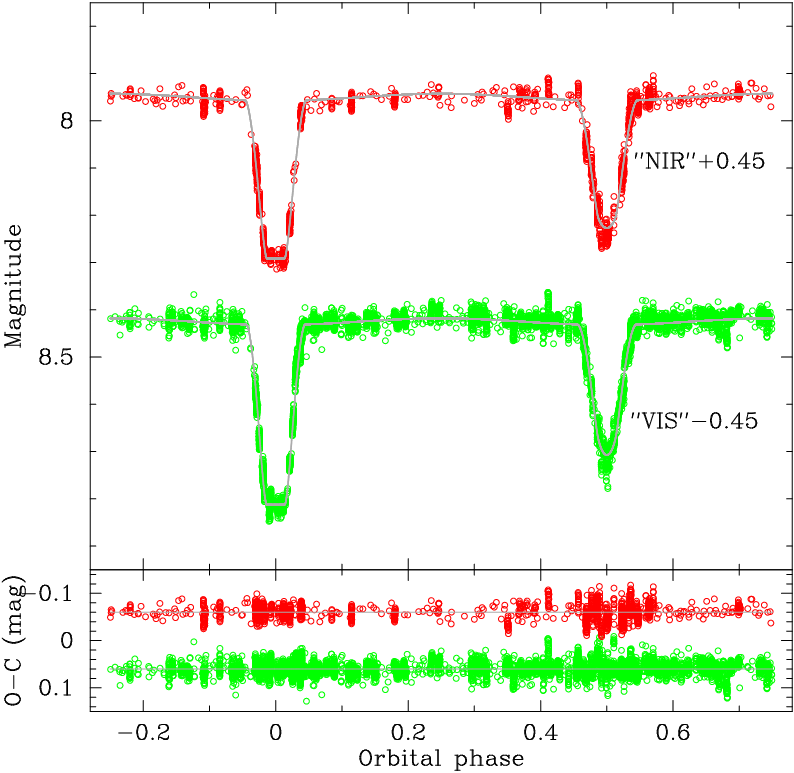}
\caption{Same as Fig.~\ref{fig_lc_jkt} but for WD models, 
and combined ''visual'' and ''near-infrared'' light curves.
Colour version of the figure available in the on-line version of the manuscript.
}\label{fig_lc_wd}
\end{figure}

\subsection{Simultaneous RV+LC analysis with WD (G-group)}\label{sec_wd}
The G-group made the binary model using all data together at the same time.
The code used in the analysis was the 2007 version of Wilson-Devinney program \citep{wil71,wil79,wil90,van07}. We simultaneously solved all light and radial velocity curves. The light curves were divided into two groups: ``visual'' -- containing all observations in ASAS $V$-band, SuperWASP and PROMPT $V$-band data and ``near-infrared'' -- containing ASAS $I$-band and PROMPT $I$-band data. Within both groups some slight shifts were done to adjust SuperWASP and PROMPT magnitude scales to ASAS magnitudes. The differences in the mean depth and width of the eclipses between different data sets are smaller than systematic effects (night-to-night variations) we noticed in the light curves. In total, the ``visual'' and ``near-infrared'' light curves contain 7121 and 1653 points, respectively. We used radial velocities derived from the Broadening Function analysis and we applied a shift of +40 \ms\ to primary's velocities to account for its blueshift. The approach to find model solution was essentially similar to method described by \cite{gra14}. The difference was that the primary's effective temperature was set as free parameter instead of secondary's one. The reason was that we estimated unique surface temperature of the secondary component from atmospheric analysis of the totality spectrum $T_2=4360\pm 80$~K.

We set [$Fe/H$]$=-0.3$ from the atmospheric analysis with \textsc{moog}. The orbital period was kept as a free parameter of a solution. We assumed circular orbit and synchronous rotation of both components. We also checked for the third light, but the fit resulted in negative values, thus we kept it fixed to zero. Logarithmic limb darkening law was used \citep{kli70}. In total we adjusted 11 parameters of the model. The model LCs are presented in Fig.~\ref{fig_lc_wd}. The resulting parameters are shown in Table~\ref{tab_wd_res}. We note that our
effective temperatures are closer to the values from the \citet{wor11} calibrations obtained by the H-group, than to their \textsc{sme} results.

\begin{table}
\centering
\caption{Results of the WD fit (G-group)}
\label{tab_wd_res}
\begin{tabular}{lcc}
\hline \hline
Parameter& Primary& Secondary \\
\hline
$P_{obs}$ (d) &\multicolumn{2}{c}{88.3865(27)} \\
$T_{0}$ (JD-2452000) & \multicolumn{2}{c}{92.034(97)} \\
$a$ ($R_{\sun}$) & \multicolumn{2}{c}{120.51(4)}\\
$q$ & \multicolumn{2}{c}{1.0004(5)} \\
$i$ (deg) &\multicolumn{2}{c}{87.68(15)}\\
$\gamma$ (km s$^{-1}$)& $ -16.19(1)$ & $-16.15(1)$\\
$\Omega^a$ & 11.629(77) & 6.323(19)\\
$r$ & 0.0941(7) & 0.1887(7) \\
$T_{\rm eff}$ (K) & 4687(5) & 4360$^b$ \\
$L_V$ & 3.607(7) & 8.956(16)\\
$L_I$ & 3.237(6) & 9.307(16)\\
$K$ (km s$^{-1}$)& 34.458(15)&  34.444(15)\\
RV $rms$ (m s$^{-1}$) & 43 & 42\\
$V$-band $rms$ (mag) &\multicolumn{2}{c}{0.008} \\
$I$-band $rms$ (mag) & \multicolumn{2}{c}{0.016} \\
DOF &\multicolumn{2}{c}{8815} \\
$\chi^2/$DOF & \multicolumn{2}{c}{0.991}\\
\hline
\end{tabular}
\\
$^a$ Dimensionless equipotential of the Roche model.\\
$^b$ Fixed.
\end{table}

\section{Physical parameters}

\subsection{G-group}
Absolute values of parameters were calculated with the WD code, assuming the same astronomical constants as in Table 5 of \cite{gra12}. The distance to the system  was derived using \citet{dib05} calibration of visual surface brightness vs. $(V\!-\!K)$ colour relation appropriate for giant stars and expressed in Johnson photometric system. We used 2MASS magnitudes from~\cite{cut03}: $J=6.492$ mag and $K=5.674$ mag and extrapolated components' light ratio in $J$- and $K$-band from the WD model $l_{21}(J)=3.26$, $l_{21}(K)=3.65$. The 2MASS magnitudes were converted onto Johnson's system using equations given by~\cite{bes88} and  \cite{car01}\footnote{\texttt{http://www.astro.caltech.edu/$\sim$jmc/2mass/v3/\\transformations/}}.  The interstellar reddening was derived from reddening maps~\citep{sch98} using normalization given by~\cite{sch11}, assuming a distance to our target (see below) and galactic dust distribution consistent with thin disc model from ~\cite{dri01}. The resulting $E(B\!-\!V)=0.121 \pm 0.016$, where error takes into account uncertainty of total extinction from \cite{sch98} maps and Milky Way's thin disc model parameters. The BC's were calculated from~\cite{alo99} calibration for given effective temperature. The derived extinction is almost equal to the extinction estimated by the H-group, which puts confidence in our approach and resulting distance of $614 \pm 18$ pc. The distance corresponds to parallax of $1.63\pm0.05$ mas.

\begin{table*}
\centering
\caption{Comparison of used approaches. For each method a main advantage and presumable source of systematic errors is given}
\label{comp}
\begin{tabular}{lp{1.7cm}p{2.2cm}p{2.2cm}p{1.7cm}p{2.2cm}p{2.2cm}}
\hline \hline
 & \multicolumn{3}{c}{\bf G-group} & \multicolumn{3}{c}{\bf H-group} \\
Analysis stage & Method& Advantages & Systematics  & Method& Advantages & Systematics \\
\hline
RVs derivation & RaVeSpAn & direct determina\-tion from BF & use of templates & tomography~\& least-squares fitting & use of disentangled spectra & initial template mismatch \\
&&&&&&\\
Atmospheric & \textsc{moog}&  well calibrated against temperature standards & use of LTE & \textsc{sme} and line depth ratios& line profiles fit\-ting, also blends &use of LTE \\
&&&&&&\\
Light curves & WD & all light curves simultaneously & fluxes from LTE models; activity & \jkt & fast; red noise accounted for & stellar activity \\
&&&&&&\\
RV curves & WD & tidal corrections included & relativ. effects not included & \textsc{v2fit} & relativistic eff- ects included & tidal corrections not included \\
&&&&&&\\
Distance & SB~--~colour relation & direct, empirical & SB calibration & \jktabs & average of various methods & calibration of the methods used \\
\hline
\end{tabular}
\end{table*}

\begin{table*}
\centering
\caption{Physical parameters of the system.}
\label{par_fi}
\begin{tabular}{lcccccc}
\hline \hline
& \multicolumn{2}{c}{\bf G-group$^a$} & \multicolumn{2}{c}{\bf H-group$^b$} & \multicolumn{2}{c}{\bf Adopted} \\
Parameter& Primary& Secondary & Primary& Secondary & Primary& Secondary \\
\hline
Spectrum & K0 III$^c$ &  K2.5 III$^c$ & K0.5 III$^d$ & K2.5-3 III$^d$& K0-0.5 III & K2.5 III \\
$M$ (M$_{\sun}$) & 1.501(2) & 1.502(2) & 1.507(3) & 1.509(3) & 1.504(4) & 1.505(4) \\
$R$ (R$_{\sun}$) & 11.34(9) & 22.74(9) & 11.31(28)& 22.50(71)& 11.33(28)& 22.62(50)\\
$\log{g}$ (cgs)  & 2.505(6) & 1.901(3) & 2.509(21)& 1.913(27)& 2.507(20)& 1.907(19)\\
$T_{\rm eff}$ (K)    & 4687(85) & 4360(80) & 4610(50)$^e$ & 4300(50)$^e$ & 4650(80) & 4330(70)\\
$L$ (L$_{\sun}$) & 55.7(4.1)&  168(12) & 51.9(3.3)&  155(12) & 53.9(3.9) & 161(13)\\
$M_{\rm bol}$ (mag)  &   0.39  & $-0.81$ &   0.46  & $-0.72$ &   0.42  & $-0.77$ \\ 
$BC_V$ (mag)     & $-0.48$ & $-0.71$ & $-0.44$ & $-0.66$ & $-0.46$ & $-0.69$ \\
$[Fe/H]$	 & $-0.25(15)$ & $-0.30(15)$ & $-0.24(12)$ & $-0.20(7)$ & \multicolumn{2}{c}{$-0.25$(10)} \\
Distance (pc) & \multicolumn{2}{c}{614(18)} & \multicolumn{2}{c}{598(18)} & \multicolumn{2}{c}{606(18)}\\
$E(B\!-\!V)$ (mag) & \multicolumn{2}{c}{0.12(2)} & \multicolumn{2}{c}{0.13(7)} & \multicolumn{2}{c}{0.13(5)} \\
\hline
\end{tabular}
\\
$^a$~Formal WD fit errors, systematics not always included;
$^b$~Systematics included;
$^c$~According to calibration by~\cite{alo99};
$^d$~According to calibration by~\cite{tok00};
$^e$~From the \textsc{sme} analysis.\\
\end{table*}

\subsection{H-group}

Absolute values of parameters and their uncertainties were calculated with the 
\jktabs code, available together with \textsc{jktebop}, assuming astronomical
constants suggested by \citet{har11}\footnote{The disparities obtained from using 
two different sets of constants are in this case negligible in comparison 
with uncertainties of derived physical parameters.}.
This simple code combines the spectroscopic and light curve solutions 
to derive a set of stellar absolute dimensions, related quantities, and distance.
We used photometry from 2MASS in $JHK$ ($J=6.492$, $H=5.674$, $K=5.674$~mag),
Tycho \citep{hog00} in $B$ (10.13~mag), and out-of-eclipse combined Johnson's 
$V$ magnitude from the \jkt solution (8.866~mag).
\jktabs calculates distances using a number of bolometric corrections for 
various filters \citep{bes98,flo96,gir02} and surface brightness-$T_{\rm eff}$ 
relations from \citet{ker04} -- 13 in our case. 
We found $E(B\!-\!V)$ for which the standard deviation of the resulting distance 
(assumed to be its uncertainty) is the lowest. Outside the given error of 
$E(B\!-\!V)$, distances differ from each other by more than 1$\sigma$.
The result -- 0.13(7)~mag -- is in a good agreement with the one
found on the basis of the secondary's $V\!-\!I$ colours -- 0.16(2)~mag.
Employing this value, and temperatures from calibrations of \citet{wor11}
-- 4710 and 4370~K -- we get a very similar distance of 604(18)~pc.

\subsection{Adopted parameters}

We combined results from analysis done by our two groups 
to derive absolute parameters. The comparison of two approaches is
presented in Tab.~\ref{comp}, and the final set of physical parameters  
in Tab.~\ref{par_fi}. As final values, we adopted straight averages of the two 
obtained by two groups. To get conservative errors, we took the average 
of the two uncertainties and added it in quadrature 
to half of the difference between the two values. When systematics were not 
included ($R$ and $\log(g)$), we assumed that they are $2\times$ the 
uncertainty given. All in all, we reached a very good precision in radii
(2.5+2.2 per cent), and one of the best estimates of stellar masses in literature
(0.27+0.27 per cent). We have also calculated the distance to the system with 
3.3 per cent error (total systematic and statistical uncertainty), which translates 
into 0.05 mas uncertainty in parallax at 606 pc (1.65 mas).
Having precisely measured distances on such scales will be important to
independently verify the results of the recently launched {\it Gaia} mission.

\section{Discussion}

\subsection{Galactic binaries with giant components}
In the on-line DEBCat catalogue\footnote{\texttt{http://www.astro.keele.ac.uk/$\sim$jkt/debcat/}}
there are only 17 system listed that have at least one star evolved and larger
than 5~R$_\odot$, and both masses and radii known with accuracy 2 per cent or better.
Of these only 3 are galactic systems (others belong to LMC or SMC) and only the 
primaries are larger than 5~R$_\odot$. These are V380~Cyg \citep[B1.5~III;][]{tka14}, 
TZ~For \citep{and91} and KIC~8410637 \citep{fra13}.
A number of other galactic systems have smaller components, although evolved from the main
sequence (AI~Phe, Andersen et al. \citeyear{and88}, He\l miniak et al. \citeyear{hel09}; 
CF~Tau, Lacy et al. \citeyear{lac12}; V432~Aur, Siviero et al. \citeyear{siv04}), or are 
much more evolved but measured less precisely (OW~Gem, Ga\l an et al. \citeyear{gal08};
$\alpha$~Aur, Torres et al. \citeyear{tor09}; ASAS~J182510-2435.5 and V1980~Sgr, 
Ratajczak et al. \citeyear{rat13}). This makes ASAS-19 the best measured, evolved
galactic binary, and a very unique object, important for studies of late 
stages of the stellar evolution. 

\subsection{Age and evolutionary status}\label{sec_age}

\begin{figure}
\centering
\includegraphics[width=\columnwidth]{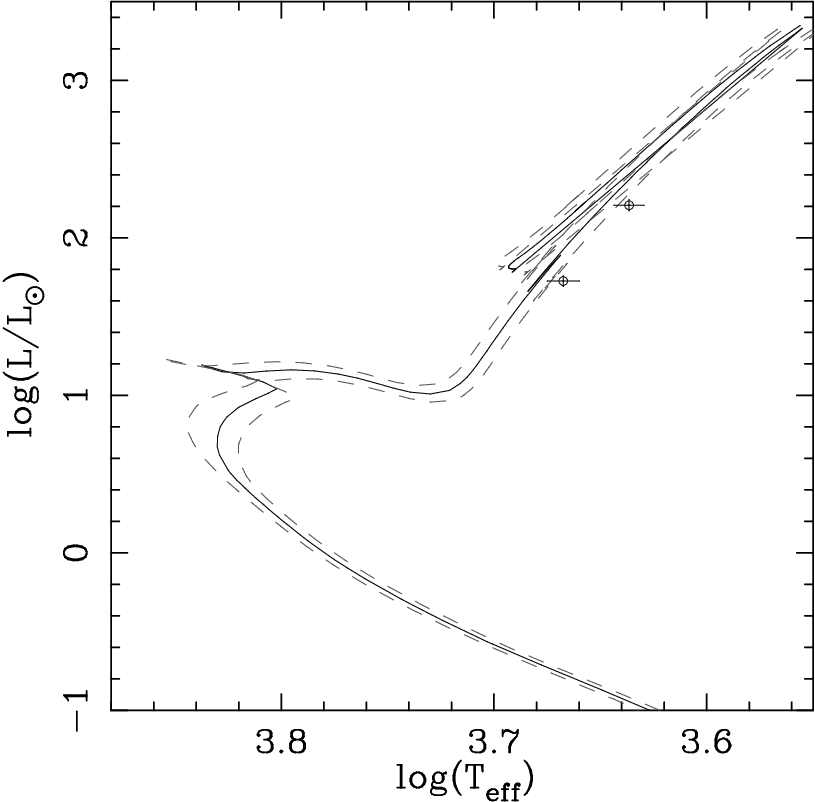}
\caption{Location of ASAS-19 on the H-R diagram. Black line is the isochrone for 
$[Fe/H]=-0.25$ and 2.38~Gyr. Two grey dashed lines are are isochrones for 
$[Fe/H]=-0.15$, 2.55~Gyr (``colder''), and $[Fe/H]=-0.35$, 2.24~Gyr (``hotter'').
}\label{fig_hr}
\end{figure}
\begin{figure}
\centering
\includegraphics[width=\columnwidth]{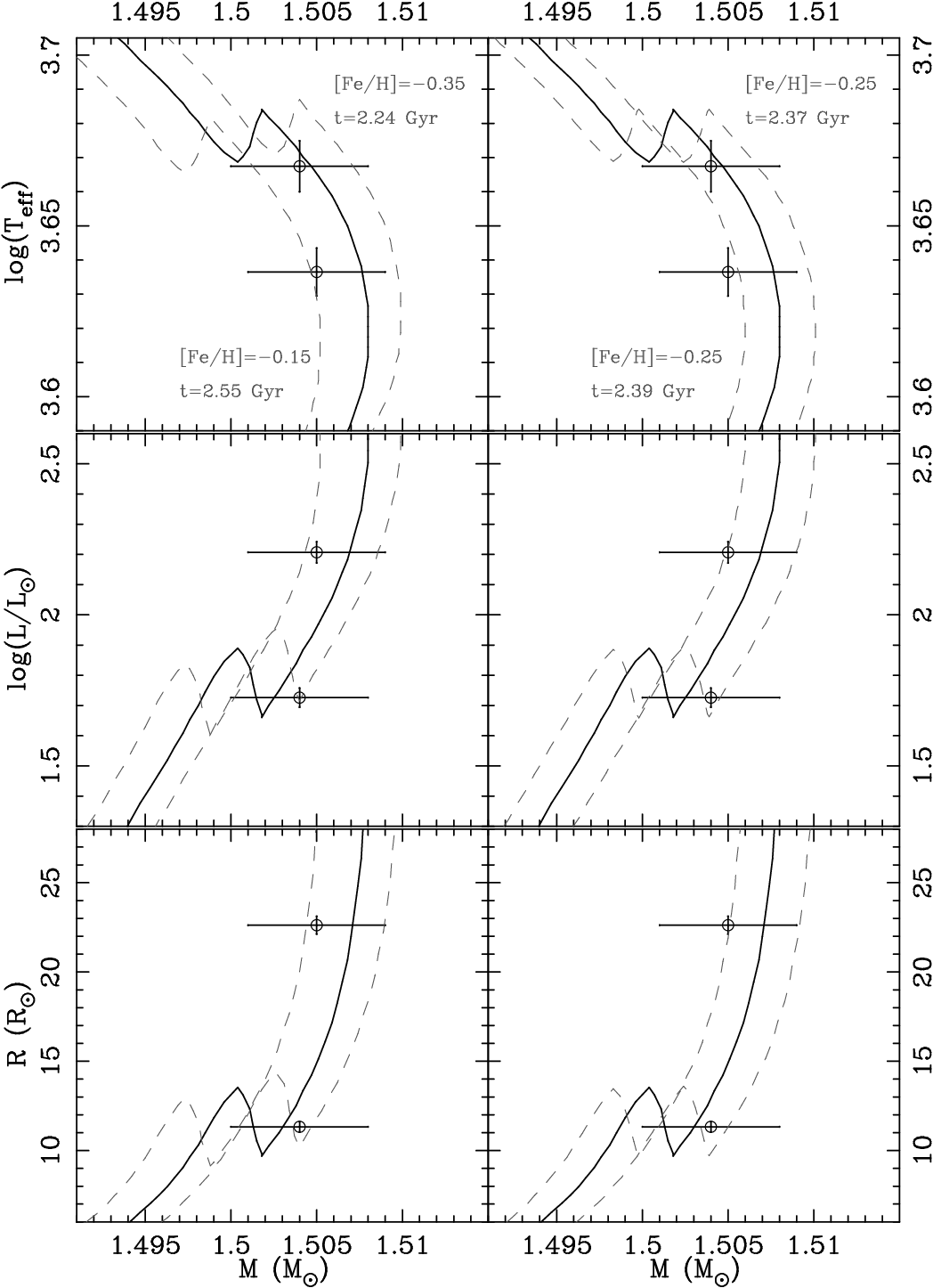}
\caption{Comparison of our final results with a 2.38~Gyr, $[Fe/H]=-0.25$ 
isochrone from the PARSEC set (black solid line). Other, marginally-fitting
isochrones are plotted in grey (dashed): on the left panels 
for $(\tau,[Fe/H])=(2.55$~Gyr,~$-0.15)$, and $(2.24$~Gyr,~$-0.35)$, 
showing the age uncertainty due to metallicity, and on the right panels for 
$(\tau,[Fe/H])=(2.37$~Gyr,~$-0.25)$, and $(2.39$~Gyr,~$-0.25)$, showing the age 
uncertainty due to mass.
}\label{fig_iso}
\end{figure}

Both stars are currently on the red giant branch, but before 
the Red Clump (Fig. \ref{fig_hr}). On this stage of evolution, 
stars of a similar mass present a wide range of radii, temperatures, 
luminosities etc., so precise mass and metallicity determination is 
crucial to constrain their age and exact evolutionary phase. 
We compared our results from Table \ref{par_fi} with stellar isochrones 
from the PAdova and tRieste Stellar Evolution Code \citep[PARSEC;][]{bre12}.
We used the value of $[Fe/H]=-0.25$, which for this set translates into 
$Z = 0.00855$, $Y = 0.2642$. We looked for the age that fits best to our
precise and direct mass measurements, and found that ASAS-19 is 
$2.38^{+0.17}_{-0.14}$~Gyr old. Most of this age uncertainty comes 
from the [$Fe/H$] determination -- for a fixed metal content, the
uncertainty coming from the mass determination is only 0.01~Gyr. 

In the Figure \ref{fig_iso} we show our results on mass vs. temperature,
luminosity and radius diagrams, together with various isochrones: 
the best-fitting (2.38~Gyr, $-0.25$~dex), two for the marginal values 
of age and metallicity that still reproduce our results within
1$\sigma$ -- 2.24~Gyr for $-0.35$, and 2.55~Gyr for $-0.15$ dex
(left, also in Figure~\ref{fig_hr}), and two more for fixed 
metallicity of $-0.25$~dex but the age of 2.37 and 2.39 Gyr (right).
One can note, that the 2.38~Gyr, $-0.25$~dex isochrone that fits the mass 
measurements best, predicts slightly hotter and more luminous stars
(Fig.~\ref{fig_hr}). This discrepancy may come from either metallicity 
or temperatures being a bit underestimated. The 2.38~Gyr, $-0.25$~dex
isochrone fits better if temperatures from the calibrations of
\citet{wor11} are used.

\subsection{Usefulness of observations during total eclipses}
The cases like ASAS-19 allow for independent verification of
indirect approaches to determine physical parameters of stars in eclipsing
binaries. It shows how the observations performed during
a total eclipse are useful for the analysis of DEBs.
Especially important was the spectrum taken when only one star was visible.
From its analysis we could independently estimate the temperature of one
of the components and metallicity of the whole system. Light curves 
alone do not constrain well the temperature scale, only the ratio of the 
two $T_{\rm eff}$-s. The common approach to light curve modelling utilizes the 
observed colour of the whole system, but it works fine only if the 
components have similar temperature or the total light is dominated
by one of them, and only if the observed colour is properly dereddened.
In our case we could securely keep one of the $T_{\rm eff}$-s fixed.
We could also calculate the observed colours of both stars, one 
directly from the photometry in the total eclipse, and the other 
from simple calculations described in Section \ref{sec_jkt_h}.
Having the multi-band photometry and the $T_{\rm eff}$ estimation from the 
spectrum, one can also calculate the $E(B\!-\!V)$ by comparing the
colours observed and predicted by colour-temperature calibrations.
For nearby systems, where the interstellar extinction is not significant, 
the observed colours would be enough to calculate the temperature 
of both components.

We have also used the totality spectrum to estimate the metallicity of the system.
This helped us to constrain the age of the binary. The well known age-metallicity
degeneration is weaker for red giants than for main sequence stars, but is still 
present. As we've found in Section \ref{sec_age}, 0.1 dex uncertainty in [$Fe/H$]
translates into 0.1 Gyr error in age. For main sequence objects it is at least
10 times more, but it would still be enough to discriminate between stars that
have just started their MS evolution, and those that are about to finish it soon.

Metallicity can also be estimated from tomographically disentangled spectra, but 
the disentangled spectra have to be correctly renormalized in order to account 
for the companion's continuum which dilutes the depth of the absorption lines.
It is relatively easy for systems showing total eclipses, as from the depth of 
this eclipse it is straightforward to calculate the contribution of each component,
and it also allows us to check if the flux ratio inferred from \todcor is 
correct. It is also possible to verify the results of decomposition by comparing the
decomposed and totality spectra, as in Figure \ref{fig_tomo}. As one can see, 
the disentangled spectra are of higher S/N, however, the approach we used (H-group) 
requires at least 8 observations in evenly-spread orbital phases.
For totally-eclipsing systems, having a single observation during the total eclipse
is less time-consuming and can give important results with less effort.
We also want to note, that the decomposition itself is easier, as for each
observed composite spectrum it is required to know only two parameters: the 
velocity difference for the component visible in totality and the flux ratio, 
both of which can be estimated separately or are easy to fit for.

Finally we want to emphasize that a high signal-to-noise spectrum taken 
during totality can also be a very good template for RV measurements of at least 
one component, as it obviously matches its $T_{\rm eff}$, $\log{g}$, [$Fe/H$] 
and turbulence velocities.

\section*{Acknowledgements}

We would like to thank the staff of the ESO La Silla observatory for 
their support during observations, and the anonymous Referee for
valuable comments and suggestions that helped to improve this work.

K.G.H. acknowledges support provided by the National Astronomical Observatory 
of Japan as Subaru Astronomical Research Fellow, and the Polish National Science 
centre grant 2011/03/N/ST9/01819.
We (D.G., B.P., G.P., P.K., K.S.) gratefully acknowledge financial support for this work
from the Polish National Science centre grant 
2013/09/B/ST9/01551 and the TEAM subsidy from the
Foundation for Polish Science (FNP).
D.G, G.P. and W.G. are supported by the BASAL Centro de Astrofisica y Tecnologias Afines (CATA) PFB-06/2007
D.G and W.G. also acknowledge support from the Millenium Institute of Astrophysics (MAS) of the Iniciativa Cientifica Milenio del Ministerio de Economia, Fomento y Turismo de Chile, project IC120009.
S.V. gratefully acknowledges the support provided by Fondecyt reg. no. 1130721.
M.K. is supported by the European Research Council Starting Grant, the
Polish National Science centre through grant 5813/B/H03/2011/40, the
Ministry of Science and Higher Education through grant W103/ERC/2011, 
and the Foundation for Polish Science through grant "Ideas for Poland".
M.R. is supported by the Polish National Science centre through grant 2011/01/N/ST9/02209.
This research was supported in part by the National Science Foundation through 
Grants 0959447, 0836187, 0707634 and 0449001, and by the European Social Fund 
and the national budget of the Republic of Poland within the framework of the 
Integrated Regional Operational Programme, Measure 2.6. Regional innovation
strategies and transfer of knowledge -- an individual project of the
Kuyavian-Pomeranian Voivodship ``Scholarships for Ph.D. students
2008/2009 -- IROP''.

We have used data from the WASP 
public archive in this research. The WASP consortium comprises 
of the University of Cambridge, Keele University, University of 
Leicester, The Open University, The Queens University Belfast, 
St. Andrews University and the Isaac Newton Group. Funding for 
WASP comes from the consortium universities and from the UKs 
Science and Technology Facilities Council.


\label{lastpage}


\begin{thebibliography}{99}
\bibitem[Alonso et al.(1999)]{alo99} Alonso A., Arribas S., Mart{\'i}nez-Roger C., 1999, A\&ASS, 140, 261
\bibitem[Andersen et al.(1988)]{and88} Andersen J., Clausen J. V., Nordstr{\"o}m B., Gustaffson G., Vandenberg D. A., 1988, A\&A, 196, 128
\bibitem[Andersen et al.(1991)]{and91} Andersen J., Clausen J. V., Nordstr{\"o}m B., Tomkin J., Mayor M., 1991, A\&A, 246, 99
\bibitem[Baines et al.(2014)]{bai14} Baines E. K., Armstrong J. T., Schmitt H. R., Benson J. A., Zavala R. T., van~Belle G. T., 2014, ApJ, 781, 90
\bibitem[Bedding et al.(2010)]{bed10} Bedding T. R. et al., 2010, ApJ, 713, 176
\bibitem[Bessell \& Brett(1988)]{bes88}  Bessell M. S., Brett J. M., 1988, PASP, 100, 1134
\bibitem[Bessell et al.(1998)]{bes98} Bessell M. S., Castelli F., Plez B., 1998, A\&A, 333, 231
\bibitem[Bressan et al.(2012)]{bre12} Bressan A., Marigo P., Girardi L., Salasnich B., Dal~Cero C., Rubele S., Nanni A., 2012, MNRAS, 427, 127
\bibitem[Carpenter(2001)]{car01} Carpenter J. M., 2001, AJ, 121, 2851 
\bibitem[Coehlo et al.(2005)]{col05} Coelho P., Barbuy B., Mel{\'e}ndez J., Schiavon R. P., Castilho B. V., 2005, A\&A, 443, 735
\bibitem[Cutri et al.(2003)]{cut03} Cutri R. M. et al., 2003, VizieR, Online Data Catalogue, 2246, 0
\bibitem[di~Benedetto(2005)]{dib05} di~Benedetto G. P., 2005, MNRAS, 357, 174 
\bibitem[Drimmel \& Spergel(2001)]{dri01} Drimmel, R. \& Spergel, D. N., 2001, ApJ, 556, 181
\bibitem[Flower(1996)]{flo96} Flower P. J., 1996, ApJ, 469, 355 
\bibitem[Frandsen et al.(2013)]{fra13} Frandsen S. et al., 2013, A\&A, 556, A138
\bibitem[Ga{\l}an et al.(2008)]{gal08} Ga{\l}an C., Miko{\l}ajewski M., Tomov T., Kolev D., Graczyk D., Majcher A., Janowski J. L., Cika{\l}a M., 2008, Observatory, 128, 298
\bibitem[Girardi et al.(2002)]{gir02} Girardi L., Bertelli G., Bressan A., Chiosi C., Groenewegen M. A. T., Marigo P., Salasnich B., Weiss A., 2002, A\&A, 391, 195
\bibitem[Graczyk et al.(2012)]{gra12} Graczyk D. et al., 2012, ApJ, 750, 140
\bibitem[Graczyk et al.(2014)]{gra14} Graczyk D. et al., 2014, ApJ, 780, 59
\bibitem[Harmanec \& Pr\v{s}a(2011)]{har11} Harmanec P., Pr\v{s}a A., 2011, PASP, 123, 976
\bibitem[He\l miniak et al.(2009)]{hel09} He\l miniak K. G., Konacki M., Ratajczak M., Muterspaugh M. W., 2009, MNRAS, 400, 969
\bibitem[He\l miniak et al.(2011)]{hel11} He\l miniak K. G. et al., 2011, A\&A, 527, A14
\bibitem[Houk(1982)]{hou82} Houk N., 1982, Michigan Catalogue of Two-dimensional Spectral Types for the HD stars. Volume 3. Declinations -40 deg to -26 deg.
\bibitem[H{\o}g et al.(2000)]{hog00} H{\o}g E. et al., 2000, A\&A, 355, L27
\bibitem[Kallinger et al.(2009)]{kal09} Kallinger T., Weiss W. W., De~Ridder J., Hekker S., Barban C., 2009, ASC, 404, 307 
\bibitem[Kervella et al.(2004)]{ker04} Kervella P., Th\'evenin F., Di~Folco E., S\'egransan D., 2004, A\&A, 426, 297 
\bibitem[Klinglesmith \& Sobieski(1970)]{kli70} Klinglesmith D. A., Sobieski S., 1970, AJ, 75, 175
\bibitem[Konacki et al.(2010)]{kon10} Konacki M., Muterspaugh M. W., Kulkarni S. R., He\l miniak, K. G., 2010, ApJ, 719, 1293
\bibitem[Lacy et al.(2012)]{lac12} Lacy C. H. S., Torres G., Claret A., 2012, AJ, 144, 167
\bibitem[Mayor et al.(2003)]{may03} Mayor M. et al., 2003, The Messenger, 114, 20
\bibitem[Marino et al.(2008)]{mar08} Marino, A. F., Villanova, S., Piotto, G., et al., 2008, A\&A, 490, 62
\bibitem[Pietrzy{\'n}ski et al.(2013)]{pie13} Pietrzy{\'n}ski G. et al., 2013, Nature, 495, 76
\bibitem[Pilecki et al.(2012)]{pil12} Pilecki B., Konorski P., G{\'o}rski M., 2012, {From Interacting Binaries to Exoplanets, IAU Symposium}, 282, 301
\bibitem[Pojma\'nski(2002)]{poj02} Pojma\'nski G., 2002, AcA, 52, 397
\bibitem[Pollacco et al.(2006)]{pol06} Pollacco D. L. et al., 2006, PASP, 118, 1407
\bibitem[Popper \& Etzel(1981)]{pop81} Popper D. M., Etzel P. B., 1981, AJ, 86, 102
\bibitem[Porter \& Woodward(2000)]{por00} Porter D. H., Woodward P. R., 2000, ApJS, 127, 159
\bibitem[Ram{\'i}rez \& Allende Prieto(2011)]{ram11} Ram{\'i}rez, I. \& Allende Prieto, C., 2011, ApJ, 743, 135
\bibitem[Ratajczak et al.(2013)]{rat13} Ratajczak M., He{\l}miniak K. G., Konacki M., Jord\'an A., 2013, MNRAS, 433, 2357
\bibitem[R\'o\.zyczka et al.(2009)]{roz09} R\'o\.zyczka M., Ka\l u\.zny J., Pietrukowicz P., Pych W., Mazur B., Catel\'an M., Thompson I. B., 2009, AcA, 59, 385
\bibitem[Rucinski(1992)]{ruc92} Rucinski S. M., 1992, AJ, 104, 1968
\bibitem[Rucinski(1999)]{ruc99} Rucinski S. M., 1999, in Hearnshaw J. B., Scarfe C.~D.,~eds, ASP Conf. Ser. Vol. 185, IAU Colloquium 170, Precise Stellar Radial Velocities. Astron. Soc. Pac., San Francisco, p. 82  
\bibitem[Schlafly \& Finkbeiner(2011)]{sch11} Schlafly, E. F. \& Finkbeiner, D. P., 2011, ApJ, 737, 103
\bibitem[Schlegel et al.(1998)]{sch98} Schlegel, D. J., Finkbeiner, D. P. \& Davis, M., 1998, ApJ, 500, 525
\bibitem[Schwarzschild(1975)]{sch75} Schwarzschild M., 1975, ApJ, 195, 137
\bibitem[Siviero et al.(2004)]{siv04} Siviero A., Munari U., Sordo R., Dallaporta S., Marrese P. M., Zwitter T., Milone E. F., 2004, A\&A, 417, 1083
\bibitem[Sneden(1973)]{sne73} Sneden C., 1973, ApJ, 184, 839
\bibitem[Southworth(2008)]{sou08} Southworth J., 2008, MNRAS, 386, 1644
\bibitem[Southworth et al.(2004a)]{sou04a} Southworth J., Maxted P. F. L., Smalley B., 2004a, MNRAS, 351, 1277
\bibitem[Southworth et al.(2004b)]{sou04b} Southworth J., Zucker S., Maxted P. F. L., Smalley B., 2004b, MNRAS, 355, 986
\bibitem[Southworth et al.(2011)]{sou11} Southworth J., Pavlovski K., Tamajo E., Smalley B., West R. G., Anderson D. R., 2011, MNRAS, 414, 3740
\bibitem[Strassmeier \& Schordan(2000)]{str00} Strassmeier K. G., Schordan P., 2000, AN, 321, 277
\bibitem[Tokunaga(2000)]{tok00} Tokunaga A.T. 2000, in Allen's Astrophysical Quantities, 4th edition, ed. A.N. Cox, Springer-Verlag (New York), p. 143.
\bibitem[Tkachenko et al.(2014)]{tka14} Tkachenko A. et al., 2014, MNRAS, 438, 3093
\bibitem[Torres et al.(2009)]{tor09} Torres G., Claret A., Young P. A., 2009, ApJ, 700, 1349
\bibitem[Torres et al.(2010)]{tor10} Torres G., Andersen J., Gimenez A., 2010, A\&ARv, 18, 67
\bibitem[Valenti \& Piskunov(1996)]{val96} Valenti J. A., Piskunov N., 1996, A\&AS, 118, 595
\bibitem[van~Hamme(1996)]{vHa96} van~Hamme W., 1996, AJ, 106, 2096
\bibitem[van Hamme \& Wilson(2007)]{van07} van Hamme W., Wilson R. E., 2007, ApJ, 661, 1129
\bibitem[Villanova et al.(2010)]{vil10} Villanova, S., Geisler, D., \& Piotto, G., 2010, ApJ, 722, 18
\bibitem[Wilson \& Devinney(1971)]{wil71} Wilson R. E., Devinney E. J., 1971, ApJ, 166, 605
\bibitem[Wilson(1979)]{wil79} Wilson R. E., 1979, ApJ, 234, 1054
\bibitem[Wilson(1990)]{wil90} Wilson R. E., 1990, ApJ, 356, 613
\bibitem[Windmiller(2010)]{win10} Windmiller G., Orosz J. A., Etzel P. B., 2010, ApJ, 712, 1003
\bibitem[Worthey \& Lee(2011)]{wor11} Worthey G., Lee H.-C., 2011, ApJS, 193, 1
\bibitem[Zucker \& Mazeh(1994)]{zuc94} Zucker S., Mazeh T., 1994, ApJ, 420, 806

\end{thebibliography}
\end{document}